\documentclass[twocolumn,english,aps,prb]{revtex4-2}
\usepackage[T1]{fontenc}
\usepackage[utf8]{inputenc}
\usepackage{color}
\usepackage[english]{babel}
\usepackage{verbatim}
\usepackage[dvipsnames]{xcolor}
\usepackage{amsbsy}
\usepackage{amstext}
\usepackage{graphicx}
\usepackage{multirow}

\usepackage[unicode=true,pdfusetitle,
 bookmarks=true,bookmarksnumbered=false,bookmarksopen=false,
 breaklinks=false,pdfborder={0 0 0},pdfborderstyle={},backref=false,colorlinks=true]{hyperref}
\hypersetup{urlcolor=blue, citecolor=blue}


\usepackage{babel}
\begin{document}
\title{Physical properties of $R$Co$_{2}$Al$_{8}$ single crystals ($R=$ La, Ce, Pr, Nd and Sm): An emerging structure-type for anisotropic Kondo lattice studies}
\author{Fernando A. Garcia$^{1,2,3}$, Sushma Kumari$^{1,2}$, Juan Schmidt$^{1,2}$,
Cris Adriano$^{2}$, Aashish Sapkota$^{1}$, Paul C. Canfield$^{1,2}$,
Rebecca Flint$^{1,2}$ and Raquel A. Ribeiro$^{1,2}$}
\affiliation{$^{1}$Ames Laboratory, U.S. DOE, Ames, Iowa 50011, USA}
\affiliation{$^{2}$Department of Physics and Astronomy,
Iowa State University, Ames, Iowa 50011, USA}
\affiliation{$^{3}$Instituto de Física, Universidade de São Paulo, São Paulo-SP,
05508-090, Brazil}
\begin{abstract}
Systematic investigations of rare-earth ($R$) based intermetallic
materials are a leading strategy to reveal the underlying mechanisms
governing a range of physical phenomena, such as the formation of
a Kondo lattice and competing electronic and magnetic anisotropies.
In this work, the magnetic, thermal and transport properties of $R$Co$_{2}$Al$_{8}$
($R=$ La, Ce, Pr, Nd and Sm) single crystals are presented. LaCo$_{2}$Al$_{8}$
is characterized as a Pauli paramagnet and transport measurements,
with the current along and perpendicular to the orthorhombic $c$-axis ($\rho_{c}$
and $\rho_{ab}$, respectively), reveal a clear electronic anisotropy,
with $\rho_{ab }\approx(4-7)\rho_{c }$ at $300$ K. We show that CeCo$_{2}$Al$_{8}$
is a Kondo-lattice for which the Kondo coherence temperature $T_{\text{K}}^{*}$, deduced from broad maximums in $\rho_{c}$ and $\rho_{ab}$  at $\approx$ 68 and 46 K, respectively, is also anisotropic. This finding is related to a possible underlying
anisotropy of the Kondo coupling in CeCo$_{2}$Al$_{8}$. The Pr
and Nd-based materials present strong easy-axis anisotropy ($c$-axis)
and antiferromagnetic (AFM) orders below $T=4.84$ K and $T=8.1$ K,
respectively. Metamagnetic transitions from this AFM to a spin-polarized
paramagnetic phase state are investigated by isothermal magnetization
measurements. The Sm-based compound is also an easy-axis AFM with
a transition at $T=21.6$ K. 
\end{abstract}
\maketitle

\section{Introduction}

The physical properties of rare-earth ($R$) based intermetallic materials
encompass a broad phenomenology ranging from, among others, heavy
fermion behavior, superconductivity and magnetism. Rich phase diagrams
are found in the case of heavy fermion materials, reflecting the coexistence
of energy scales competing for the material ground state. The formation
of the heavy fermion state is due to the local hybridization
between the rare-earth-derived $4f$ and conduction electrons, and is more often realized in Ce- and Yb-based materials.
This hybridization leads to the formation of a Kondo lattice, characterized
by a many-body coherence temperature $T_{\text{K}}^{*}$, below which
the localized and itinerant states are entangled in a liquid of heavy
carriers \citep{hewson_kondo_1993,coleman_quantum_2005,steglich_foundations_2016,yang_emerging_2022,coleman_introduction_2016}.

The key to understanding heavy fermions is to disentangle the contributions
from distinct degrees of freedom, lattice and electronic, which is
usually accomplished by systematic investigations of the physical
properties of a particular series. Indeed, for a given structural
type, it is paramount to seek for examples of non-moment bearing (La-,
Lu- and Y-based materials ), pure local moment (Pr-, Nd-, Gd-Tm-based
materials ) and hybridizing (Ce- and Yb-based materials and, sometimes,
Pr-, Sm- and Eu-based materials) $R$-based intermetallics. This need
for a comprehensive set of physical properties is well exemplified
by material series like $R$Ni$_{2}$Ge$_{2}$ \citep{budko_anisotropy_1999},
$R$Ni$_{2}$B$_{2}$C \citep{bhatnagar_electrical_1997,canfield_rni2b2c_1997,fisher_anisotropic_1997},
$R$AgSb$_{2}$ \citep{petrovic_anisotropic_2003,myers_systematic_1999},
$R$PtBi \citep{canfield_magnetism_1991,mun_rptbi_2022} and $R$$M$$_{2}$Zn$_{20}$
\citep{torikachvili_six_2007}.

An ongoing topic of investigation in this field is the interaction
between the Kondo lattice and the magnetic anisotropies. Indeed, the
local hybridizations leading to the Kondo lattice are set by large
energy scales that may compete with the crystalline electric field
(CEF) determined anisotropy. This is well illustrated by the ``hard-axis'' ordering phenomenology, 
where magnetic order develops along the (CEF determined) hard-axis of magnetization in Kondo materials
\citep{andrade_competing_2014,hafner_kondo-lattice_2019,kwasigroch_magnetic_2022,kornjaca_distinct_2024}. 
Moreover, the subject motivates the pursuit of first-principles theoretical approaches to understanding
CEF effects and the search for new anisotropic Ce-based materials \citep{xia_machine_2025,lee_importance_2025}. 

Recently, attention has been devoted to 1-2-8 $R$-based materials
adopting the CaCo$_{2}$Al$_{8}$-type orthohombic crystal structure
(space group $Pbam$). In this structure, the ($R$) cations
are organized along chains in the $c$-axis, with the $R$-$R$ distances
being smaller along the $c$-axis than the distances observed when
the structure is projected onto the $ab$-plane.  In particular, some
gallides ($R$$M_{2}$Ga$_{8}$, where $M$ is a transition metal)
have shown interesting physics, as magnetic quantum critical behavior
observed in NdFe$_{2}$Ga$_{8}$ \citep{wang_neutron_2022,wang_quantum_2021}
and an axial Kondo lattice in CeCo$_{2}$Ga$_{8}$ \citep{zou_abnormal_2024,zheng_uniaxial_2022,cheng_realization_2019,wang_heavy_2017}.
In the later case, an axial Kondo chain is evidenced by a strong anisotropy
of the resistivity ($\rho$), which shows a broad maximum (about $17$
K, characterizing $T_{\text{K}}^{*}$ for this material), only for
measurements along the $c$-axis ($\rho_{c}$). For measurements along
$a$ and $b$, $\rho$ keeps going up at low temperatures in a way
that is reminiscent of the single ion Kondo effect (incoherent Kondo
scattering). This phenomenology suggests that the underlying $4f$-conduction
electrons hybridization is anisotropic in this material. The CeCo$_{2}$Ga$_{8}$
magnetic properties are also anisotropic, with the $c$-axis identified
as the easy-axis. The $c$-axis was also identified as the easy-axis
for PrFe$_{2}$Ga$_{8}$ \citep{wang_single-crystal_2022} and PrRu$_{2}$Ga$_{8}$
\citep{xiao_strong_2023}. As for the vast majority of the $R$-based
materials, the magnetic anisotropy is likely set by crystalline electric
field (CEF) effects \citep{cheng_realization_2019}.

The $R$$M_{2}$Ga$_{8}$ materials thus display anisotropic electronic
and magnetic properties. CeCo$_{2}$Ga$_{8}$, in particular, showcases
an example of coexisting hybridization and CEF determined anisotropies.
A good number of $R$-based aluminides ($R$$M_{2}$Al$_{8}$) also
adopt the CaCo$_{2}$Al$_{8}$-type orthohombic structure. Early explorations
of these materials focused on samples in polycrystalline form \citep{ghosh_strongly_2012,nair_magnetic_2016,nair_pr-magnetism_2017}
until recently, when the growth of single crystals from an Al-rich
ternary composition was reported \citep{watkins-curry_strategic_2015,treadwell_investigation_2015}.
Our paper is dedicated to the physical properties of $R$Co$_{2}$Al$_{8}$
($R=$ La, Ce, Pr, Nd and Sm) single crystals. 

We start showing that LaCo$_{2}$Al$_{8}$ is a simple Pauli paramagnet
which display a clear electronic anisotropy, that is characterized
by a large anisotropy in resistivity ($\rho$) measurements. Indeed,
at $T=300$ K, we found that $\rho_{ab}/\rho_{c} \approx 4-7$, where
$\rho_{c}$ and $\rho_{ab}$ are, respectively, the resistivity measured
with the current along and perpendicular to the orthorhombic, $c$-axis. We then investigated CeCo$_{2}$Al$_{8}$.

Previously, the formation of a Kondo lattice in CeCo$_{2}$Al$_{8}$
was deduced based upon the observation of a broad maximum in $\rho(T)$
measurements of polycrystalline samples \citep{ghosh_strongly_2012}
but this was not observed in CeCo$_{2}$Al$_{8}$ single crystals
\citep{watkins-curry_strategic_2015}. The status of CeCo$_{2}$Al$_{8}$
as a Kondo lattice thus requires clarification. Here, we show that
CeCo$_{2}$Al$_{8}$ is an anisotropic Kondo lattice system, for which
$T_{\text{K}}^{*}$, deduced from broad maximums in $\rho_{c}$ and
$\rho_{ab}$ ($T_{\text{K,}c}^{*}$ and $T_{\text{K,}ab}^{*}$ , respectively),
assume different values, with $T_{\text{K,}c}^{*}\approx68$ K and
$T_{\text{K,}ab}^{*}\approx46$ K. This finding is examined in the
light of the proposed Kondo chain in CeCo$_{2}$Ga$_{8}$.

We then investigate PrCo$_{2}$Al$_{8}$ and NdCo$_{2}$Al$_{8}$.
PrCo$_{2}$Al$_{8}$ was shown to be a strong easy-axis antiferromagnetic
(AFM) material \citep{xiao_strong_2023}. Our experiments confirm
this property of the Pr-based material and adds that NdCo$_{2}$Al$_{8}$
is also a strong easy-axis AFM material. Moreover, we show that the
PrCo$_{2}$Al$_{8}$ undergoes two consecutive AFM transitions. For
both materials, applying a magnetic field ($H$) along the $c$-axis
induces metamagnetic transitions from the AFM state to a high field
polarized paramagnetic phase state at relatively low $H$. The $T$
vs. $H$ phase diagrams for the samples are constructed based upon
magnetization ($M(H,T)$) measurements. The Sm-based material was
previously described as a Pauli paramagnet (where Sm would assume
a $2+$ valence) \citep{watkins-curry_strategic_2015}. Here, we show
that it is a moment bearing compound with an AFM transition, $T_N \approx 22$ K. 

\section{Methods}

Motivated by previous work \citep{watkins-curry_strategic_2015},
$R$Co$_{2}$Al$_{8}$ single crystals ($R=$ La, Ce, Pr, Nd and Sm)
were obtained from an Al-rich ternary composition ($1:2:20$). A total
amount of about $3.5$ g of reactants were weighted and placed in
Canfield crucible set (CCS) \citep{canfield_new_2020,noauthor_canfield_nodate}.
Guided by the methods explained elsewhere \citep{canfield_new_2020},
the heat treatment was slightly modified to obtain phase pure single
crystals: from room temperature ($RT$), the reactants were heated
for $6$ hours to $T=1180$ $^{\circ}$C. The mixture was left at
this temperature for $24$ hours and then slowly cooled down to $T=900$
$^{\circ}$C at a rate of $2$ $^{\circ}$$\text{C}/\text{h}$, at which point
the growth was taken out of the furnace and spun to segregate the
flux from the crystals. The samples thus obtained were rod-like in
shape with typical size about $2$-$3$ mm in length with a cross
section of about $1\times1$ mm. Larger crystals, about $1$ cm in
length (see inset in figure $1$$(b)$), could be obtained by either
of the two methods: $i)$ cooling down the melt from $T=1180$
\textdegree C at a slower rate of $1$ \textdegree $\text{C}/\text{h}$ or by $ii)$ growing the crystals in a two step process; first cooling the melt to 1025 \textdegree C and decanting, which promotes the removal of any oxide slags that can serve as unwanted nucleation sites. This makes the solution more pure.  Then, secondly, taking the materials that were decanted at 1025 \textdegree C, resealing and heating to 1100 \textdegree C and cooling to 900 \textdegree C over 150 hours.

We have observed that when attempting to grow $R$Co$_{2}$Al$_{8}$ single crystals with heavier lanthanides (Gd - Tm), we obtained a new phase: $R_{2}$Co$_{6}$Al$_{19}$ whose preparation and detailed physical properties
will be the subject of a separate report. Indeed, as exemplified by
phases containing Ce-Co-Al, Pr-Co-Al and Nd-Co-Al \citep{moze_crystal_1997,tung_specific_2004,tougait_crystal_2006},
many ternary $R$-Co-Al phases are possible. 

Pieces of the obtained samples were selected, crushed and passed to a 90$ \mu$m sieved to obtain a fine powder to perform powder x-ray diffraction (XRD) experiments.
A commercial table top Rigaku Miniflex X-ray diffractometer was employed
for the XRD experiments. The GSAS2 software \citep{toby_gsas-ii:_2013}
was used to analyze the powder profile and check the crystallographic
phase. The crystal orientation was checked by Laue diffraction experiments.
We could clearly distinguish the crystallographic $c$-axis (the axis
along the rod length) but not the $a$ and $b$ axes, due to the formation
of twins in the $ab$ plane. As a consequence of this, all anisotropic data is given only as parallel to the identified $c$-axis and perpendicular to the identified $c$-axis. This means that we are not able to study any possible in-plane anisotropy in this system.

Magnetization $(M$) measurements (as function of $H$ and $T$, $M(H)$
and $M(T)$, respectively) were performed in commercial SQUID magnetometers
from Quantum Design. Measurements were performed down to $T=1.8$
K and fields as high as $\mu_{0}H=7$ T. Experiments were performed
with the applied field either along the $c$-axis (denoted $H$ $||$
$c$) or perpendicular to the $c$-axis directions (denoted $H$ $||$ $ab$).
For all samples, zero field cooling (ZFC) and field cooling (FC) $M(T)$
measurements were performed and no notable differences were found.
In this paper, we present the ZFC measurements. 

Heat capacity, $C_{p}(T)$, and electrical transport measurements
were performed in a commercial Dynacool Physical Property Measurement
System (PPMS) from Quantum Design. In the case of the La- and Ce-based
system, the electrical transport was characterized adopting two configurations,
namely the current parallel and perpendicular to the $c$-axis (denoted
$\rho_{c}$ and $\rho_{ab}$, respectively). For the other $R$-based materials, the electrical transport
was characterized only for the current parallel to the $c$-axis. 

\section{Results and Discussion}

\subsection{Crystal structure and LaCo$_{2}$Al$_{8}$ physical properties}

\begin{figure}
\begin{centering}
\includegraphics[width=0.95\columnwidth]{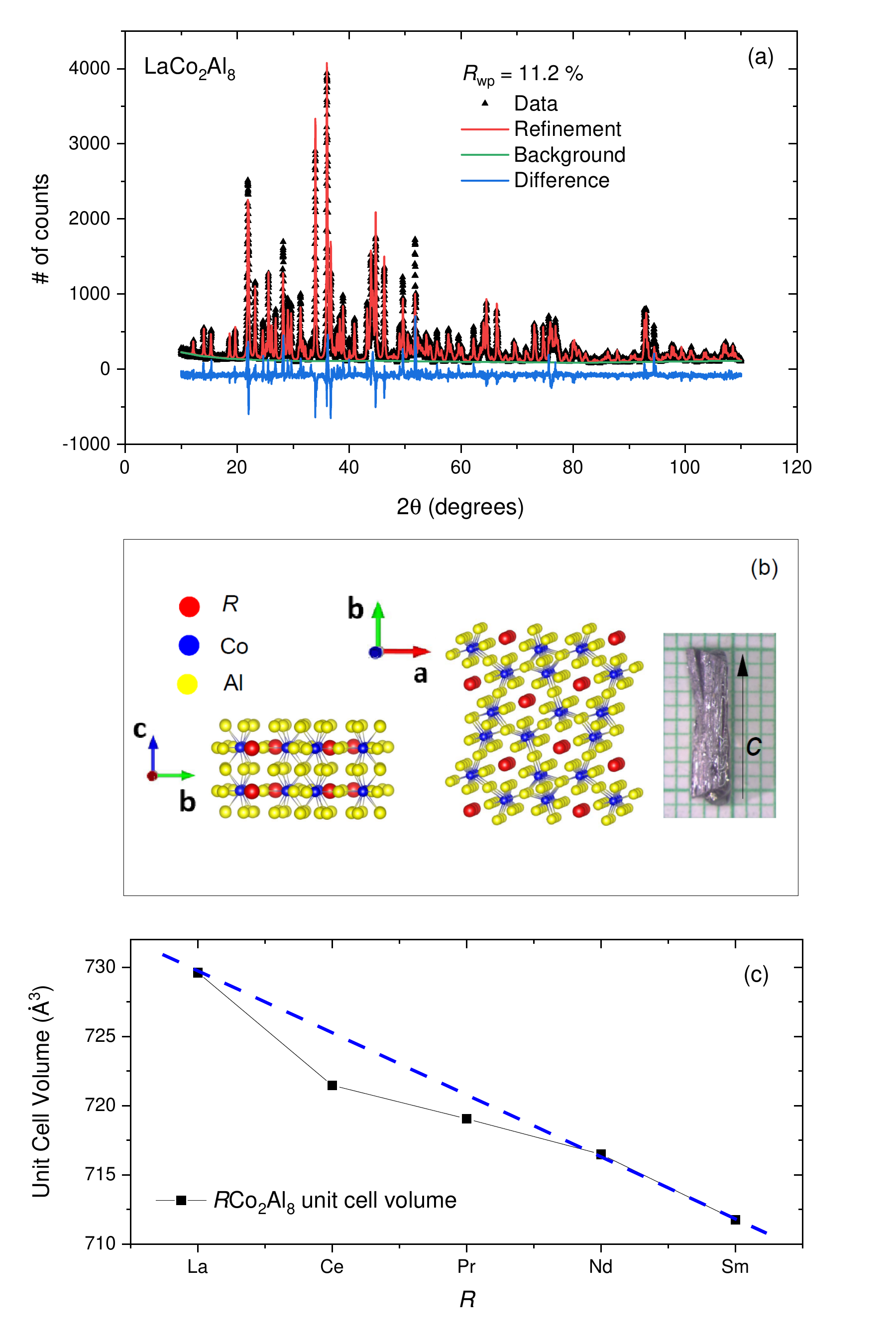}
\par\end{centering}
\caption{$(a)$ Black-squares: LaCo$_{2}$Al$_{8}$ powder XRD experimental
data, respective powder profile refinement (thick red line), background
fitting (thick green line) and difference pattern (thick blue line)
obtained with a $R_{\text{WP }}\approx11.2$ \%. $(b)$ Structural
model of the CaCo$_{2}$Al$_{8}$-type orthohombic crystal structure
(space group $Pbam$). $R$ atoms are represented by red spheres,
Co atoms by blue spheres and the Al atoms yellow spheres. Some key
structural features are highlighted by the model: $i)$ the $R$-chains
along the $c$-axis and the skewed triangular lattice in $ab$-plane
formed by the $R$ atoms $ii)$ the Co coordination structure by the
Al atoms and $iii)$ the cage-like structure around the $R$ sites.
In the inset, we show a large CeCo$_{2}$Al$_{8}$ single crystal
(see methods), about $1$ cm in lenght along the $c$-axis. $(c)$ Unit
cell volume of the $R$Co$_{2}$Al$_{8}$ materials as function of
$R$ , as obtained by Rietveld refinement of the powder XRD data (typical
$R_{\text{WP }}\approx11.0-12.0$ \%, error bars are of the order
of the point size). 
\protect\label{fig:RE_structure}}
\end{figure}

Powder XRD data were collected for all $R$Co$_{2}$Al$_{8}$ ($R=$
La, Ce, Pr, Nd and Sm) materials. The resulting diffraction patterns
were refined assuming the previously reported data from single crystal
diffraction experiments \citep{watkins-curry_strategic_2015} (space
group $Pbam$), providing a good description of experimental data.
The lattice constants assume typical values as $c\sim4.0$ $\text{\AA}$,
$a\sim12.0$ $\text{\AA}$ and $b\sim14$.0 $\text{\AA}$. Figure
\ref{fig:RE_structure}$(a)$ present the LaCo$_{2}$Al$_{8}$ powder
XRD data (black circles) and refinement results (red and blue thick
lines). The quality of the data analysis is representative of the
series. 

\begin{figure}
\begin{centering}
\includegraphics[width=0.95\columnwidth]{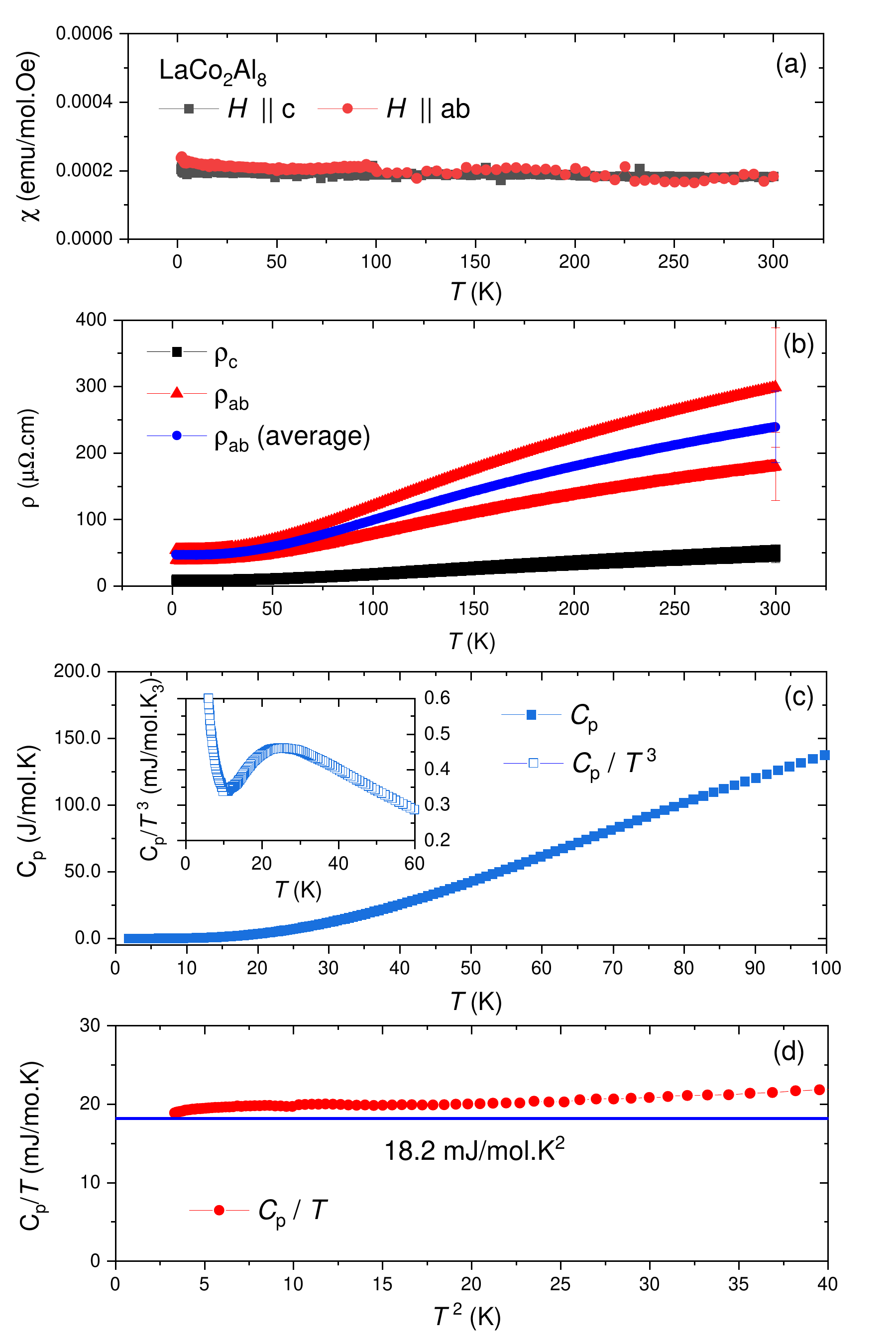}
\par\end{centering}
\caption{LaCo$_{2}$Al$_{8}$ physical properties. $(a)$ $\text{\ensuremath{\chi}}$
as a function of $T$ obtained for $\mu_{0}H=0.1$ T, for $H$ along
the $c$-axis ($H$ $||$ $c$) and in the $ab$ plane ($H$ $||$
$ab$). $(b)$ $\rho_{c }$ and $\rho_{ab}$ as a function of $T$,
in the interval $2<T<300$ K. $(c)$ $C_{p}$ as function of $T$. The
inset presents $C_{p}/T^{3}$ as function of $T$, which displays
a broad peak about $T=25$ K (see main text). (d) $C_{p}/T$ as a function of $T^{2}$ in a limited $T-$range. A reference
line (solid blue line) marks the extrapolated value of of $C_{p}/T$
as $T\rightarrow0$ . \protect\label{fig:LaCoAl_physicalprop}}
\end{figure}

In figure \ref{fig:RE_structure}$(b)$, the refined structure is
represented. Focusing first on the structural features related to
the $R$ atoms (red spheres), it can be observed that along the $c$-axis
the $R$ atoms are organized along somewhat isolated chains. The model also highlights the cage-like
structure formed around the $R$ site, which is reminiscent of what
is found in intermetallic cage-like materials. Considering
the structure projected in the $ab$ plane, the sublattice formed
by the $R$ atoms is a motif containing skewed triangles for which
AFM interactions might be somewhat frustrated. The Co-Al polyhedra show that there
is no direct $R$-Co contacts, and thus the role of the hybridization
between the $R$-derived orbitals and the Co-derived $3d$ orbitals
is negligible. The Co atoms are coordinated by $9$ Al atoms and the
typical Co-Al distances are less than the sum of the Co and Al metallic
radii, suggesting a significant covalent character of the Co-Al bond.
This feature weakens the Co magnetism, as is demonstrated by the Pauli-paramagnetism of the LaCo$_{2}$Al$_{8}$ crystal shown in Fig. \ref{fig:LaCoAl_physicalprop}$(a)$. Thus, $R$Co$_{2}$Al$_{8}$ magnetic properties are dominated by those of the trivalent $R$-ions. 

Figure \ref{fig:RE_structure}$(c)$ shows the unit cell volume, obtained
from the powder XRD refinement along the series, and the results are
in good agreement with the single crystal refinement \citep{watkins-curry_strategic_2015}. Whereas there is a clear Lanthanide contraction, the unit cell for CeCo$_{2}$Al$_{8}$ appears to be somewhat lower than would be expected, suggesting some possible degree of tetravalent nature, or perhaps more accurately, some degree of hybridization.

Assuming the partitioning of the electronic (localized $4f$
electrons plus conduction electrons) and lattice degrees of freedom
of the other $R$Co$_{2}$Al$_{8}$ materials, LaCo$_{2}$Al$_{8}$
can thus be rationalized as a suitable reference material to obtain
the $4f$-electron contributions to physical properties of the other
$R$-based materials. We thus investigate the LaCo$_{2}$Al$_{8}$ magnetization, electrical transport and heat capacity ($C_{p}$). Turning to measurements of physical properties, Fig.\ref{fig:LaCoAl_physicalprop}$(a)$ presents data on the LaCo$_{2}$Al$_{8}$ single crystalline sample. The LaCo$_{2}$Al$_{8}$ magnetization $M$ as function of
$T$, obtained with an applied field $\mu_{0}H$ of $0.1$ T, is presented
in figure \ref{fig:LaCoAl_physicalprop}$(a)$. We define the system
susceptibility, $\chi$, as $M/H$. The results are nearly $T$-independent,
as expected for a Pauli paramagnet, and the observed value is $\approx2\times10^{-4}$
emu/mol.Oe. 

Resistivity ($\rho$) measurements (current $||$ $c$-axis and $||$
$ab$-plane,  denoted as $\rho_{\text{c }}$ and $\rho_{\text{ab }}$,
respectively) are presented in \ref{fig:LaCoAl_physicalprop}$(b)$
for $2<T<300$ K. Metallic behavior with residual resistivities of
about $\rho_{c }\approx 8 - 10$ $\mu\Omega$.cm and $\rho_{ab }\approx 40 - 55$
$\mu\Omega$.cm are observed at $T=2$ K. Similarly, a factor in between $\approx 4 - 7$ is observed for $\rho_{ab}/\rho_{c}$ at $T=300$ K, characterizing a strong electronic anisotropy.

Room temperature resistivity ratio ($RRR$) parameters between  $\approx5-6$ were obtained, similar to what we observed for the other $R$-based materials. Resistivity
measurements were performed for $3$ samples ($\rho_{\text{c}}$)
and $2$ samples ($\rho_{ab}$), each with its own error bar. In the former case, typical dimensions
for the distance between the electrodes ($L$) were $\approx0.8-1.1$
mm and the cross section area ($A$) were in between $\approx0.18-0.20$
mm$^{2}$. In the latter case, measurements were executed with values
in between $L\approx0.4-0.6$ mm and $A\approx0.25-0.30$ mm$^{2}$.  For $\rho_{ab}$, the actual measurements 
differ by a factor of about $\approx1.6$ and we also present the average between the two curves to be inspected in reference with the estimated error bars. 
 
The $C_{p}$ data as function of $T$ are presented in Fig. \ref{fig:LaCoAl_physicalprop}(c).
From now on, we shall denote by $C_{p}^{R}$ the heat capacity of
the $R$-based material. No sign of a phase transition was observed
down to $T=1.81$ K, as previously reported \citep{ghosh_strongly_2012}.
The $C_{p}^{\text{La}}$ data assume an almost constant value in the
low-$T$ region. Indeed, in the simplest approximations, it is expected
that $C_{p}$ at sufficient low-$T$ should be modeled as $C_{p}/T=\text{\ensuremath{\gamma}}+\beta T^{2}$,
where $\gamma$ is the Sommerfeld coefficient, connected with the
conduction electrons and $\beta$ is a constant connected with the
Debye (phonon) contributions to $C_{p}$. The adequacy
of this model is investigated in the upper inset of the figure wherein
we plot $C_{p}^{\text{La}}/T^{3}$ as a function of $T$. 

A broad peak at about $25$ K is observed. Such contribution is reminiscent of a low frequency non-dispersive phonon,  usually paraphrased as a ``rattling mode'' of the $R$ atoms inside a cage-like structure. This vibration is connected with good potential for thermoelectric applications in cage-like materials such as skutterudites and clathrates
\citep{hermann_einstein_2005,tian_atomistic_2008,snyder_complex_2008,mao_advances_2018}. 
These modes are modeled, in first approximation, by an Einstein contribution to the heat capacity which contains a single parameter: the Einstein temperature $\Theta_{\text{E}}$. A peak about $25$ K in the heat capacity data is associated with $\Theta_{\text{E}}\approx125$ K. Whereas this number stands out as a rather low-frequency optical vibration, vibrations as low as $\approx1$ GHz ($\approx5$ K) were already characterized for rattlers in skutterudite materials \citep{garcia_coexisting_2009}.

Figure \ref{fig:LaCoAl_physicalprop}$(d)$ plots $C/T$ versus $T^2$ for $0 < T^2 < 30$ K. Whereas the data for $T^2 < 25$ K is well approximated by the Debye model with $\gamma^{La} \approx 18.2(5) $ mJ/mol.K$^2$ and $\beta^{La} \approx 0.45(2)$ mJ/mol.K$^2$, for $T^2 > 25$ this is not the case. This is most likely associated with the low-lying Einstein mode discussed above.

The $\gamma^{\text{La}}$ was obtained per mol of formula unit and
it is then close to $\approx1.5$ mJ/mol.K$^{2}$ per atom, as expected
for simple metals, i.e., our estimate for $\gamma^{\text{La}}$ does
not suggest enhancement of the quasiparticle mass. Indeed, the ratio
between the Pauli-like response and $\gamma^{\text{La}}$ ($2\times10^{-4}/\gamma^{\text{La}}$
emu.K$^{2}$/mJ) is $\approx1.24\times10^{5}$ (emu.K$^{2}$/mJ )
close to $1.37148\times10^{5}$ (emu.K$^{2}$/mJ) deduced for a simple
free electron gas \citep{ashcroft_solid_1976}. Overall, LaCo$_{2}$Al$_{8}$
is a Pauli paramagnet that can be adopted as a ``non-magnetic''
reference material for the $R$Co$_{2}$Al$_{8}$ ($R=$ Ce, Pr, Nd
and Sm) compounds. This said, we should note three important points. First, similar Einsten modes should exist for the other ( $R$ = Ce, Pr, Nd, and Sm) members of the $R$Co$_2$Al$_8$ family. We do not know the exact value of their Einstein temperatures, but given the lanthanide contraction, it will probably shift in some monotonic manner. Second, even with this uncertainty, the LaCo$_2$Al$_8$ $C_p$ data remain a good non-magnetic analogue for subtraction from the other $R$Co$_2$Al$_8$ materials because even with a possible small shift in Einstein temperature, the total entropy change over a wide temperature range will be well modeled by the LaCo$_2$Al$_8$. Third, as such, we will not be able to use the typical low-$T$  heat capacity model to extract values for $\gamma$ and $\beta$ for the other $R$Co$_2$Al$_8$ compounds.

\subsection{The Kondo lattice CeCo$_{2}$Al$_{8}$}

\begin{figure}
\begin{centering}
\includegraphics[width=0.95\columnwidth]{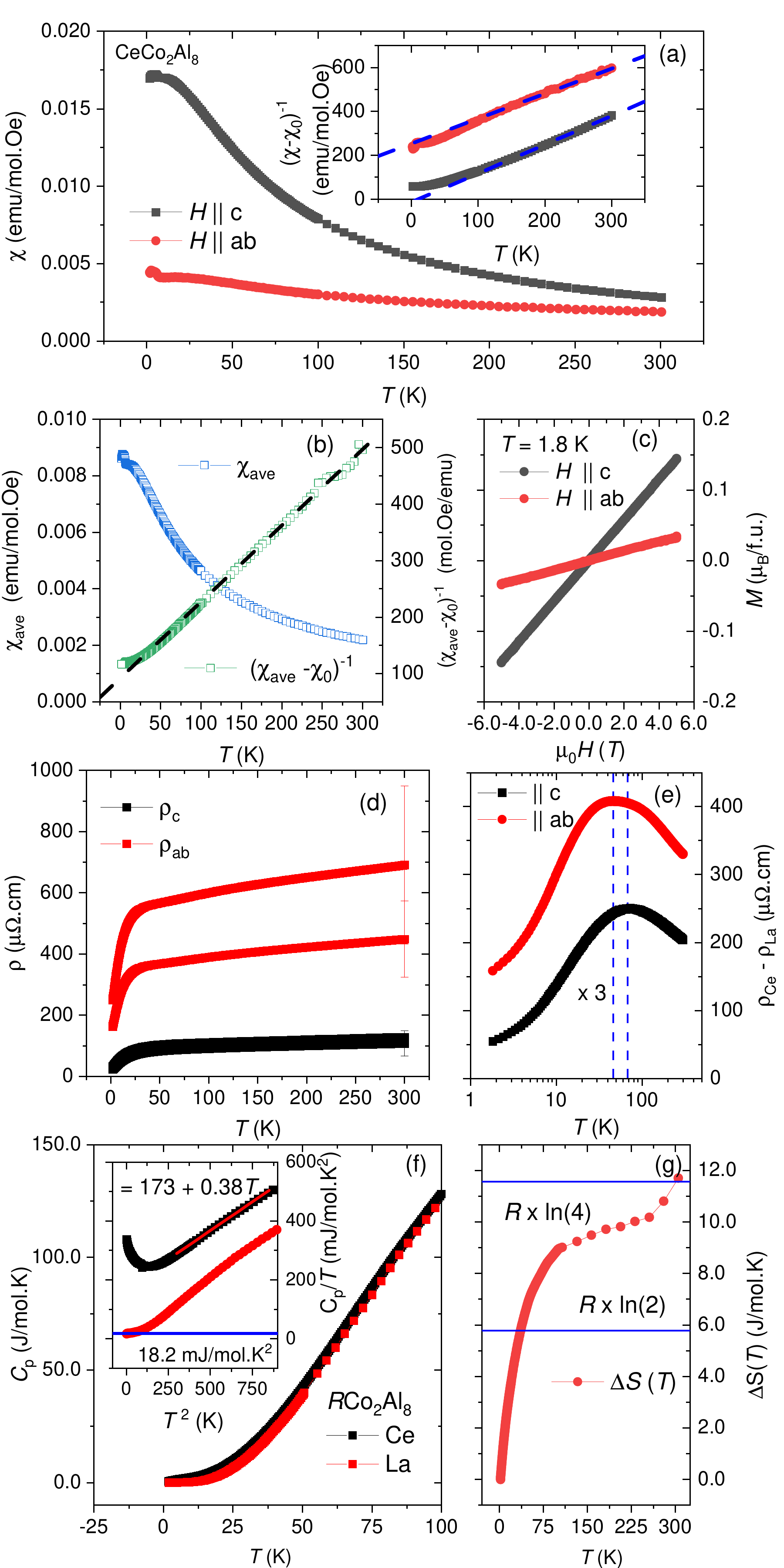}
\par\end{centering}
\caption{CeCo$_{2}$Al$_{8}$ physical properties. $(a)$ The CeCo$_{2}$Al$_{8}$
$\chi_{c}$ and $\chi_{ab}$ . In the inset, we present the inverse
of $(\chi_{ab} -\chi_{ab_0})$ and $(\chi_{c}-\chi_{c_0})$. The dashed
blue lines represent the CW fitting (obtained for $T>150$ K) of the
data. $(b)$ Respectively on the left and right axis, $\chi_{\text{ave}}^{\text{Ce}}$and
$(\chi_{\text{ave}}^{\text{Ce}}-\chi_{0})^{-1}$ are presented. The
dashed black line represents the inverse CW fitting of the data (obtained
for $T>150$ K). $(c)$ Isothermal magnetizations obtained for $H\text{ }||\text{ }c$
and $H\text{ }||\text{ }ab$ at $T=1.8$ K. $(d)$ 
CeCo$_{2}$Al$_{8}$ $\rho_{c}$ and $\rho_{ab}$ as a function
of $T$ for a number of single crystals. $(e)$ The Ce $4$f contribution to resistivity ($\rho_{\text{4f,Ce}}=\rho_{\text{Ce}}-\text{\ensuremath{\rho_{\text{La}}}})$
as a function of $T$ in a log scale, for the two different directions.
The vertical blue line marks the maximums of $\rho_{\text{4f,Ce}}$
which we adopt as an estimate to $T_{\text{K},c}^{*}=68$ K and $T_{\text{K},ab}^{*}=46$
K. Note that average values of the respective resistivity curves were used.$(f)$ CeCo$_{2}$Al$_{8}$ and LaCo$_{2}$Al$_{8}$ heat capacity. The inset shows in detail the $C_{p}$ vs. $T^{2}$  at low-$T$. $(g)$The magnetic
entropy contribution ($\Delta S$) of CeCo$_{2}$Al$_{8}$ and some
reference values (solid blue lines). \protect\label{fig:CeCoAl_physprop}}
\end{figure}

In figures \ref{fig:CeCoAl_physprop}$(a)$-$(c)$ the magnetic properties of CeCo$_{2}$Al$_{8}$ are presented. Magnetization measurements were made for $\mu_{0}H=0.1$ T, for $H\text{ }||\text{ }c$
and $H\text{ }||\text{ }ab$. In \ref{fig:CeCoAl_physprop}$(a)$,
$\chi$ is presented and the magnetic response is clearly anisotropic,
with the $c$-axis being the easy axis of magnetization. As aforementioned,
twins were observed in the $ab$ plane and therefore measured quantities for which $H\text{ }||\text{ }ab$
are, to a certain approximation, averages between results expected
with $H$ $||$ $a$ or $b$. From now on we identify quantities
obtained in measurements with $H$ either parallel to $c$ or to $ab$
by their respective subscripts. When necessary, we shall also adopt
a label $R$ to denote a quantity associated to the $R$-based material. 

A Curie-Weiss (CW) like behavior is observed for $\chi_{c}$ down
to $T\approx15$ K, below which a tendency for saturation is observed.
This is reminiscent of the formation of an enhanced Pauli-like response
of coherent heavy carriers (contributed by the Ce $4f$ states) in
Kondo-lattice systems. Along the $ab$-plane, $\chi_{\text{ab}}$
has a weaker $T$-dependence and it also tends to saturate at low$-T$.
In the inset of \ref{fig:CeCoAl_physprop}$(a)$, we present the
inverse of $\chi_{c}$ and $\chi_{ab}$ subtracted by $\chi_{0}=2\times10^{-4}$
emu/mol.Oe which is the LaCo$_{2}$Al$_{8}$ Pauli-like response.
We then fit the data (in the $T>150$ K region) to a inverse CW expression
to obtain the Curie-Weiss constants $\theta_{c}^{\text{Ce}}=-1.6(5)$
K and $\theta_{ab}^{\text{Ce}}=-267(4)$ K. 

To estimate the effective moments and the energy scale of the interactions,
we adopt a CW description of the polycrystalline average of $\chi$,
denoted by $\chi_{\text{ave}}$.  Assuming that $\chi_{ab}$ is an average
between $\chi_{a}$ and $\chi_{b}$ , we define $\chi_{\text{ave}}$ as $\chi_{\text{ave}}=(\text{\ensuremath{\chi_{c}+2\text{\ensuremath{\chi_{ab}}})/3}}.$
In figure \ref{fig:CeCoAl_physprop}$(b)$, we present $\chi_{\text{ave}}$
(left axis) and $(\chi_{\text{ave }}-\chi_{0})^{-1}$ (right axis).
We performed the CW fitting of $(\chi_{\text{ave }}-\chi_{0})^{-1}$
(for $T>150$ K) and obtained $\mu_{\text{eff}}^{\text{Ce}}=2.43(1)$
$\mu_{B}$ and $\theta_{\text{CW}}^{\text{Ce}}=-68(2)$. The value
of $\mu_{\text{eff}}^{\text{Ce}}$ corresponds to $0.96$ of the full
value expected for the Ce$^{3+}$ cations ($2.57$ $\mu_{B}$), suggesting
that the Ce derived moments are well localized in the $T>150$ K region.
Being negative, $\theta_{\text{CW}}^{\text{Ce}}$ suggests AFM interactions.

For both directions, there is a clear upturn in $\chi$ at low-$T$,
that may suggest a paramagnetic impurity in the sample or the onset
of magnetic order. This is further investigated by isothermal magnetization
measurements at $T=1.81$ K , $H\text{ }||\text{ }c$ and $H\text{ }||\text{ }ab$,
presented in the inset of Fig.\ref{fig:CeCoAl_physprop}$(a)$.
The results clearly suggest that the system is not magnetically ordered
at this temperature.

Motivated by the anisotropy of CeCo$_{2}$Ga$_{8}$ electronic properties
and the proposal of an axial Kondo lattice in the material, we measured
$\rho^{\text{Ce}}$ with the current parallel and perpendicular to
the $c$-axis. In figure \ref{fig:CeCoAl_physprop}$(d)$, $\rho_{c}$
and $\rho_{ab}$ for the  Ce-based material are presented. As is observed,
$\rho^{\text{Ce }}$is also marked by a significant anisotropy and
by broad maximums suggesting the formation of a Kondo lattice. At
$T=300$ K, $\rho_{ab}/\rho_{c}$ is about $5$. We applied the same procedure used for the La samples to the Ce samples. The resistivity was measured in both directions across multiple single crystals as shown in the figure. 

As a way estimate the contribution of the Ce derived $4f$ states
to the CeCo$_{2}$Al$_{8}$ transport properties, we show in \ref{fig:CeCoAl_physprop}$(e)$
$\rho_{\text{4f}}^{\text{Ce}}=\rho^{\text{Ce}} -\rho^{\text{La}}$
. For better comparison, we magnify the $\rho_{c}$ data multiplying
it by a factor of $3$ as indicated in the figure. The $\rho_{\text{4f}}^{\text{Ce}}$ curves were obtained from the averages 
of the $\rho^{\text{Ce}}$ and $\rho^{\text{La}}$ presented in figures \ref{fig:CeCoAl_physprop}$(d)$ and 
\ref{fig:LaCoAl_physicalprop}$(b)$, respectively.   Broad resistivity
maximums are observed at clearly distinct positions, based upon which
we define anisotropic Kondo coherence temperatures ($T_{\text{K}}^{*}$)
of about $T_{\text{K,}c}^{*}\approx 68$ K and $T_{\text{K,}ab}^{*}\approx 46$
K. Our findings suggest the emergency of anisotropic heavy quasi particles
due to formation of a Kondo lattice in CeCo$_{2}$Al$_{8}$. 
The $RRR$ obtained for CeCo$_{2}$Al$_{8}$ is about $\approx5$,
similar to the LaCo$_{2}$Al$_{8}$ case (see figure\ref{fig:LaCoAl_physicalprop}$(b)$). By inspection, one observes that $\rho_{\text{Ce,ab}}$ is about $5$
times $\rho_{\text{Ce,c}}$ at $T=300$ K. 

Electronic correlations and the presence of heavy carriers are further
investigated in figures \ref{fig:CeCoAl_physprop}$(f)$-$(g)$, where
we present the CeCo$_{2}$Al$_{8}$ specific heat data and analysis. In figure
\ref{fig:CeCoAl_physprop}$(f)$, the CeCo$_{2}$Al$_{8}$ $C_{p}$
data is presented alongside those of LaCo$_{2}$Al$_{8}$. No
sign of a phase transition is observed down to $T=1.81$ K, as previously
reported \citep{ghosh_strongly_2012}. In comparison with the La-based
material, a systematic and clear entropy excess in the Ce-based system is observed,
in particular in the low-$T$ region. In the inset, $C_{p}/T$ data
are presented as function of $T^{2}$, for both materials. A clear
upturn, reminiscent of a Kondo system, possibly close to quantum criticality,
is observed in $C_{p}^{\text{Ce}}$ at low-$T$ as previously observed
in the case of polycrystalline samples \citep{ghosh_strongly_2012}. We can estimate the low temperature $\gamma^{Ce}$ as being between 173(3) mJ/mol.K$^{2}$ (from the extrapolation of the linear part of $C_{p}/T$ vs. $T^{2}$ to zero) and $337(5)$ mJ/mol.K$^{2}$ (from the lowest temperature value measured). As such, CeCo$_{2}$Al$_{8}$ is a moderate heavy Kondo lattice system.

Having in mind the caveat related with the presence of the low-lying
optical phonon in LaCo$_{2}$Al$_{8}$, we examine an estimate for
the CeCo$_{2}$Al$_{8}$ magnetic heat capacity, which we obtain from
$C_{\text{mag}}^{\text{Ce}}=C_{p}^{\text{Ce}}-C_{p}^{\text{La}}$.
We then integrate $C_{\text{mag}}^{\text{Ce}}/T$ to calculate the
magnetic contribution to the CeCo$_{2}$Al$_{8}$ entropy variation
$\Delta S$ up to $300$ K. This is presented in figure \ref{fig:CeCoAl_physprop}$(f)$.
As observed, $\Delta S$ only shows a tendency for saturation at high-$T$
and the saturation value is about $R\times ln(4)$ only close to $300$
K. If we take one of the conventional estimates of the thermodynamic Kondo temperature ($T_K$), namely when magnetic entropy reaches $R\times ln(2)$, we can estimate $T_K\approx36$ K.

For an alternate estimate of $T_K$, one can inspect the inset of \ref{fig:CeCoAl_physprop}$(e)$ and conclude that $C_{\text{mag}}^{\text{Ce}}$ is underestimated for
$T<1.81$ K, because of the $C_{p}^{\text{Ce}}$ upturn in this region.
It is then likely that an entropy amounting to $R\times ln(4)$ is
recovered for $T<300$ K. Thus, by adopting the generalized relation $\gamma^{\text{Ce}}T_{\text{K}}=R\times ln(4)$
\citep{torikachvili_six_2007}  one finds $T_{\text{K}}\approx70$
K, in good agreement with $T_{\text{K,}c}^{*}$. This expression assumes that
two low-lying CF doublets hybridize to form the Kondo state.

From our experiments and analysis, one can deduce the formation of
a coherent Kondo lattice featuring moderate heavy quasi-particles
in CeCo$_{2}$Al$_{8}$. The electronic and magnetic properties of
the system are anisotropic and, most strikingly, the deduced Kondo coherence
temperatures are anisotropic, with a difference of about $22$ K between
$T_{\text{K,}c}^{*}$ and $T_{\text{K,}ab}^{*}$. Our findings support
a scenario wherein Kondo coherent scattering is first achieved along
the $c$-axis whereas in a relatively broad $T$-interval the Kondo
scattering in the $ab$ plane remains incoherent, reminiscent of the
single ion Kondo effect. Moreover, in view of our results for LaCo$_{2}$Al$_{8}$,
the Kondo anisotropy in CeCo$_{2}$Al$_{8}$ is likely an effect that
derives from the conduction electrons, not necessarily related with
the local hibridizations. 

There is no clear \emph{qualitative} difference between CeCo$_{2}$Al$_{8}$
and CeCo$_{2}$Ga$_{8}$. The materials are isoelectronic and the
structural parameters are in close similarity, which renders similar
Ce-Ga and Ce-Al distances. Sample quality, as determined from $RRR$,
is also similar. Moreover, the resistivity anisotropy at $300$ K
is also comparable \citep{wang_heavy_2017,zheng_uniaxial_2022}. In
CeCo$_{2}$Ga$_{8}$, $\rho_{c}$ peaks about $17$ K whereas no coherence
peak is observed for either $\rho_{a}$ or $\rho_{b}$ down to $2$
K. This, and results from optical spectroscopy experiments, supports
the proposal of an axial Kondo chain in this material \citep{cheng_realization_2019,wang_heavy_2017}.
In view of our findings for CeCo$_{2}$Al$_{8}$, it is possible that
a coherence peak in $\rho_{a}$ and $\rho_{b}$ may exist also in
CeCo$_{2}$Ga$_{8}$, but is delayed to lower temperatures. The synthesis
of LaCo$_{2}$Ga$_{8}$ single crystal and the investigation of its
transport properties is certainly highly desirable to understand how
analogous the situation is. 

\subsection{PrCo$_{2}$Al$_{8}$ and NdCo$_{2}$Al$_{8}$}

\begin{figure}
\begin{centering}
\includegraphics[width=0.95\columnwidth]{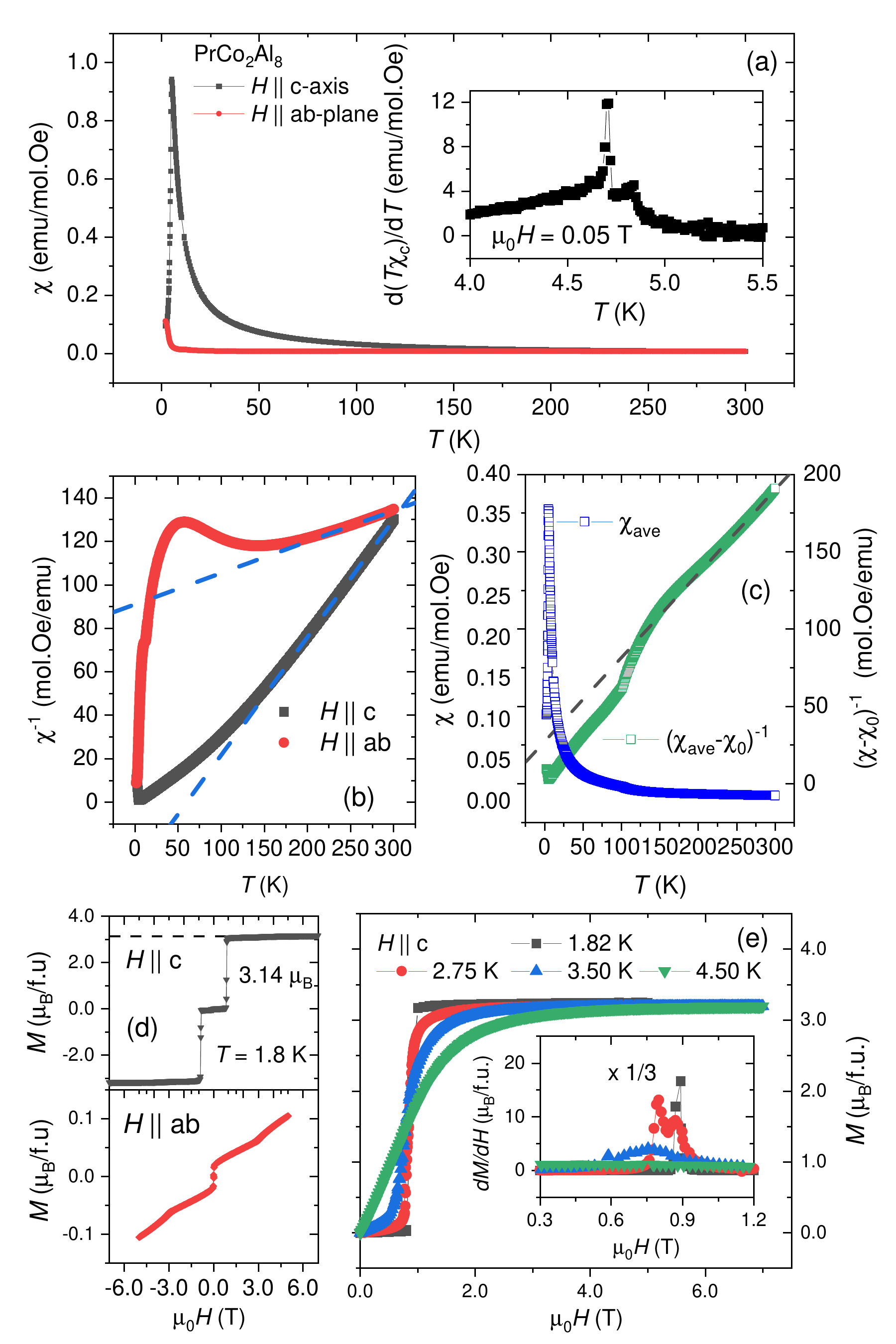}
\par\end{centering}
\caption{PrCo$_{2}$Al$_{8}$ magnetic properties. $(a)$ $\text{\ensuremath{\chi}}$
data obtained for $\mu_{0}H=0.1$ T with $H\text{ }||\text{ }c$ and
$H\text{ }||\text{ }ab$. Strong easy axis anisotropy is observed.
The measurement with $H$ along the $c$-axis suggest an AFM transition
at $T_{N}=4.84$ K . The inset shows in detail the behavior of $d(T\chi)/dT$
close to the transition temperature and two transitions are observed
($T_{N1}=4.84$ K and $T_{N2}=4.71$ K). $(b)$ The inverse of $\chi_{a}$
and $\chi_{ab}$ subtracted by $\chi_{0}$ . The dashed blue lines
represent the CW fitting (obtained for $T>225$ K) of the data. $(b)$
Respectively on the left and right axis, $\chi_{\text{ave}}$and $(\chi_{\text{ave}}-\chi_{0})^{-1}$
are presented. The solid black line represent the inverse CW fitting
of the data ($T>225$ K). $(c)$ Respectively on the left and right
axis, $\chi_{\text{ave}}$ and $(\chi_{\text{ave}}-\chi_{0})^{-1}$
data are presented (see main text for a definition of $\chi_{0}$).
The solid blue line represent the inverse CW fitting of the data ($T>225$
K) . $(d)$ Isothermal magnetization curves ($T=1.81$ K) obtained
with $H\text{ }||\text{ }c$ (upper panel) and $H\text{ }||\text{ }ab$
(lower panel). $(e)$ Representative isothermal magnetizations for
distinct temperatures obtained with $H\text{ }||\text{ }c$. The data
show clear metamagnetic transitions from a low field AFM state to
a high field spin polarized PM phase. The inset shows the $dM/d\mu_{0}H$
derivatives from which we determine the critical field for the field-induced
metamagnetic transition. The inset also shows that at some intermediate
$T$ two inflections are observed in $dM/d\mu_{0}H$. \protect\label{fig:PrCoAl-resultsA}}
\end{figure}

We now inspect, in turn, the Pr- and Nd-based materials. We start
with the former. $\chi$ measurements (figure \ref{fig:PrCoAl-resultsA}$(a)$)
display our key findings about this material: it is a strong easy-axis
magnet presenting AFM order. From the $H\text{ }||\text{ }c$ data,
a Neel temperature ($T_{N}$) of $T_{N}=4.84$ K can be deduced. In
the inset, this subject is investigated in detail. We present high
statistics measurements for $d(T\chi_{c})/dT$ about $T_{N}=4.84$
K, which adds some complexity to our findings. The data display two
putative AFM transitions, which we shall call AFM phase $I$ and $II$,
taking place at $T_{N1}=4.84$ K and $T_{N2}=4.71$ K, respectively.
Multiple magnetic transitions were also observed in other orthohombic
Pr-Co-Al ternaries, such as PrCoAl$_{4}$ \citep{schenck_multiple_2005,tung_specific_2004,schobinger-papamantellos_magnetic_2001}. 

In figure \ref{fig:PrCoAl-resultsA}$(b)$ we analyze the inverse
of $\chi_{c}$ and $\chi_{ab}$. As recently reported for this same
material \citep{xiao_strong_2023} and other Pr-based materials adopting
the CaCo$_{2}$Al$_{8}$-type structure \citep{wang_single-crystal_2022,nair_pr-magnetism_2017},
$\chi_{ab}^{-1}$ deviates strongly from the expected  CW behavior.
This deviation was attributed to excited CF levels lying about this
energy scale. In figure \ref{fig:PrCoAl-resultsA}$(a)$, close inspection
shows that indeed $\chi_{ab}$ has a broad peak-like structure (a
``belly''), suggesting that only data in the high-$T$ region should
be considered for CW fittings. We then perform the inverse CW fitting
for $T>225$ K obtaining $\theta_{c}=59(3)$ K and $\theta_{ab}=-631(10)$
K.

In figure \ref{fig:PrCoAl-resultsA}$(c)$, left axis, we present
$\chi_{\text{ave}}$. To estimate $\mu_{\text{eff}}^{\text{Pr}}$
and $\theta_{\text{CW}}^{\text{Pr}}$, we first noticed that $1/\chi_{\text{ave}}$
contains a clear deviation from linear behavior. This is most likely a manifestation of the fact that we are not performing a proper polycrystalline average but rather using $\chi_{ab}$ as a proxy for the, average, in-plane susceptibility. The fact that we still see a CEF feature in our polycrystalline average suggests that there is actually sizable in-plane magnetic anisotropy. We thus adopted a CW fitting only in the $T>225$ K region and use the fitting to extract
a constant paramagnetic contribution $\chi_{0}$ which, due to a Vlan
Vleck term, is not the same as the LaCo$_{2}$Al$_{8}$ contribution.
Then, the linear CW fitting of the $(\chi_{\text{ave}}-\chi_{0})^{-1}$
data was performed again in the restricted $T>225$ K region. This
is shown in the right axis of figure \ref{fig:PrCoAl-resultsA}$(c)$
and the fitting is represented by the black dashed line. Based on
this procedure, $\mu_{\text{eff}}^{\text{Pr}}=3.82(5)$ $\mu_{B}$
and $\theta_{\text{CW}}^{\text{Pr}}=-47(2)$ K are estimated. The
effective moments are very close to the theoretical value for Pr$^{3+}$
cations ($3.57$ $\mu_{B}$) but is an overestimation. From $\theta_{\text{CW}}^{\text{Pr}}$
, a large frustration parameter $f=|\theta_{\text{CW}} / T_{\text{N}}|\approx10$
is deduced. The $\mu_{\text{eff}}^{\text{Pr}}$ value may indicate
some contribution from Co, that may become magnetically polarized
due to the Pr presence. It has to be noted, though, given the fact that the polycrystalline average data so clearly violates CW-behavior, our values of $\theta_{\text{CW}}^{\text{Pr}}$ and $\mu_{\text{eff}}^{\text{Pr}}$ should be viewed as more qualitative than quantitative.

The properties of the AFM state are further investigated by isothermal
magnetization measurements. Results at $T=1.81$ K, for $H\text{ }||\text{ }c$
(upper panel) and $H\text{ }||\text{ }ab$ (lower panel) are presented
in figure \ref{fig:PrCoAl-resultsA}$(d)$. A metamagnetic transition
from a low field AFM state to a high field spin-polarized paramagnetic
(PM) state is observed at about $\approx0.9$ T for $H\text{ }||\text{ }c$.
The saturation value is  $\approx3.14$ $\mu_{\text{B}}/$f.u.
very close to what is expected for a $\left\Vert4\right\rangle $
singlet CF ground state. Indeed, in the case of Pr$^{3+}$ cations, which are non-Kramers ($J=4$), the degeneracy of the CF levels can be completely removed. In the case of PrCo$_{2}$Al$_{8}$, in view of the low Pr point group symmetry  ($m2$), the CF scheme should feature only singlets and doublets and possibly pseudo-doublets, which denote two nearby singlets.

For $H\text{ }||\text{ }ab$, a spin rotation
transition is observed and induced by a relatively small field. By
comparing the values of the magnetization at $1.81$ K and $\text{\ensuremath{\mu_{0}H=5}}$T
for the two field directions, the magnetic anisotropy in PrCo$_{2}$Al$_{8}$
is about $\approx32$.

\begin{figure}
\begin{centering}
\includegraphics[width=0.95\columnwidth]{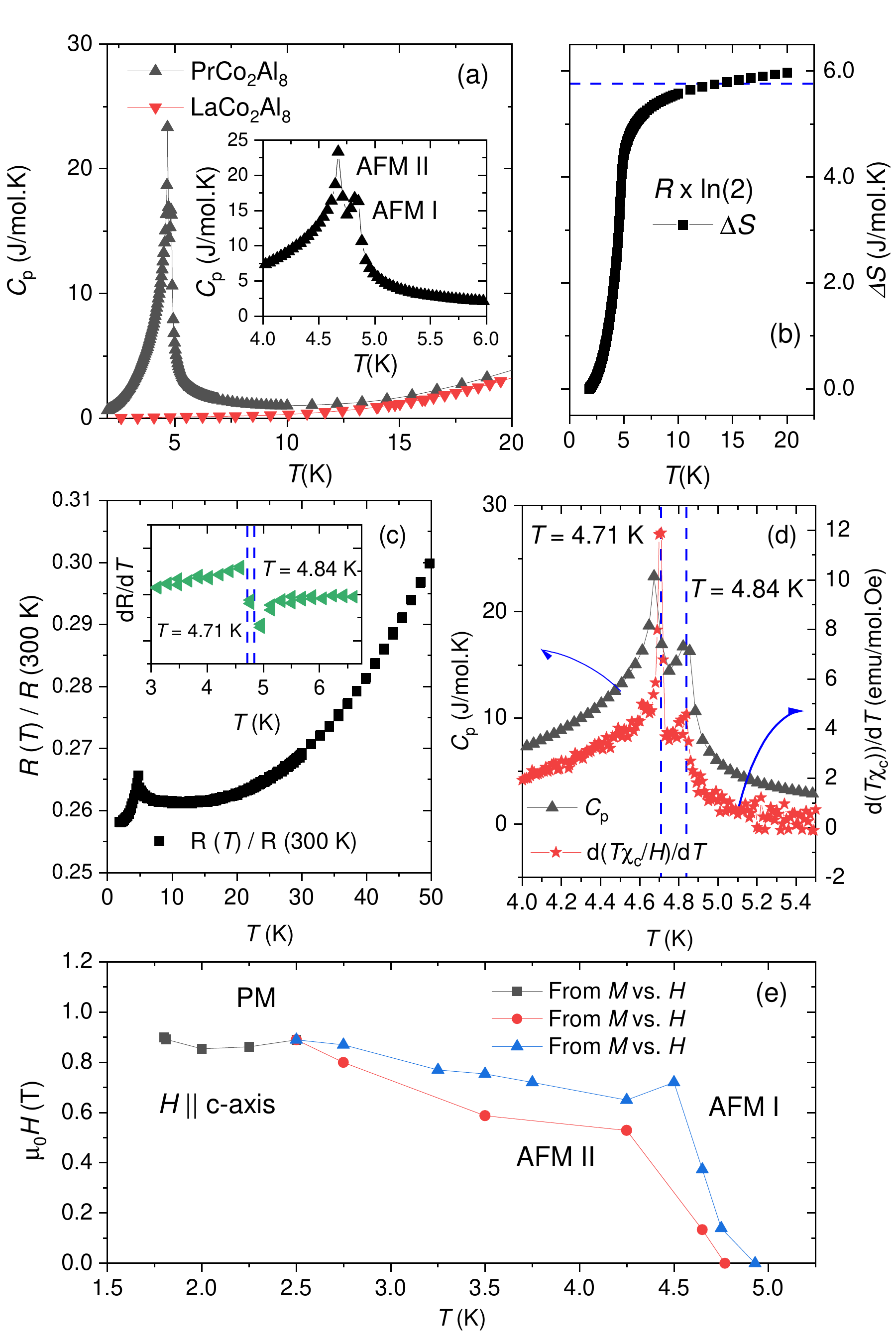}
\par\end{centering}
\caption{PrCo$_{2}$Al$_{8}$ thermal and transport properties. $(a)$ PrCo$_{2}$Al$_{8}$
$C_{p}$ (black triangles) showing the AFM transitions. The inset
provide detail data in the vicinity of $T_{N}$ and show the two consecutive
transitions which are also observed in the $\chi$ measurements. The
LaCo$_{2}$Al$_{8}$ $C_{p}$ is presented as reference. $(b)$ PrCo$_{2}$Al$_{8}$
magnetic entropy variation (see main text). $(c)$ Temperature
dependence of the normalized resistance $R(T)/R(300\text{ K})$. The
inset shows the resistance derivative to $T_{N_{1,2}}$. $(d)$
Comparison between $C_{p}$ and $d(T\chi_{c})/dT$ showcasing the
two AFM transitions as determined from independent measurements. $(e)$
PrCo$_{2}$Al$_{8}$ magnetic phase diagram ($H\text{ }||\text{ }c$)
deduced from $M(H)$ measurements. \protect\label{fig:PrCoAl-resultsB}}
\end{figure}

In figure \ref{fig:PrCoAl-resultsA}$(e)$, we explore the temperature dependence
of the metamagnetic transition, presenting some representative isothermal
magnetization curves, obtained for $H\text{ }||\text{ }c$, in the
interval $1.81<T<5$ K. In the inset, we show the field derivative
of the isothermal magnetization ($dM/d\mu_{0}H$). We adopt that a
peak in the derivative identify the critical field at which the metamagnetic
transition takes place. As can be observed, above the base temperature
(for instance for $T=2.75$ K, red circles in \ref{fig:PrCoAl-resultsA}$(e)$),
the saturation moment is reached after two peaks of the derivative.
We associate the low-field peak with a transition from AFM $II$ to
AFM $I$ and the second peak with a transition to the high field spin
polarized PM phase. For low-$T$ measurements ($T\leq2.5$ K), only
one critical field is observed. We thus conclude that at a certain $T$ and $H$, the transitions to the AFM $I$ and $II$ phases coalesce in a single transition line (see Fig. \ref{fig:PrCoAl-resultsB}$(f)$). 

The PrCo$_{2}$Al$_{8}$ $C_{p}$ (denoted $C_{p}^{\text{Pr}}$) is
presented in figure \ref{fig:PrCoAl-resultsB}$(a)$ alongside the
LaCo$_{2}$Al$_{8}$ reference data. The inset shows $C_{p}^{\text{Pr}}$
in detail the vicinity of the values of the putative AFM transition
temperatures. It clearly shows the two transitions at $T_{N1}=4.84$
K and $T_{N2}=4.71$ K. 

We investigate the total entropy associated with the magnetic ordering in 
\ref{fig:PrCoAl-resultsB}$(b)$. In all cases,  the magnetic entropy is estimated by adopting $C_{p}^{\text{La}}$ plus a renormalization factor which accounts the mass difference between La and the relevant $R$ atom. This renormalization factor is $\Theta^{\text{R}}_{\text{D}}/\Theta^{\text{La}}_{\text{D}}$ where  $\Theta^{\text{R}}_{\text{D}}$ is the Debye temperature determined from  $C_{p}^{\text{R}}$. The magnetic heat capacity is then obtained as $C_{p}^{\text{R}}\times\Theta^{\text{R}}_{\text{D}}/\Theta^{\text{La}}_{\text{D}}-C_{p}^{\text{La}}$ \citep{bouvier_specific_1991}.  In the PrCo$_{2}$Al$_{8}$ case, $\Theta^{\text{Pr}}_{\text{D}}$  and $\Theta^{\text{La}}_{\text{D}}$ were estimated from two methods: assuming the validity of the $T^{3}$ Debye law at low-$T$ and by fitting a Debye function plus an optical contribution in some selected $T$-intervals (to test the fitting stability) between $15$ K and $75$ K. The renormalization factors thus obtained ranged from $1.04(2)$  and $1.006(4)$. In view of the results, we adopted $C_{p}^{\text{La}}$ as a suitable phonon reference in this case (i.e. the renormalization factor is assumed as $\approx 1$).

The Pr magnetic heat capacity was then integrated to find the entropy recovered up to $T=25$ K, as shown in figure \ref{fig:PrCoAl-resultsB}$(b)$. The value at $\approx5$ K is already close to $R\times ln(2)$. This result, together with the $M$vs.$H$ curves (Fig.\ref{fig:PrCoAl-resultsA}$(d)$), suggests the ordering of a CF pseudo-doublet, associated with all the degrees of freedom in the sample.

In figure \ref{fig:PrCoAl-resultsB}$(c)$, we present the Pr-based
material resistance as function of temperature, $R(T).$ The data
are normalized by the value of $R(T)$ measured at $T=300$ K. The
measurements were performed with the current along the $c$-axis.
A $RRR\approx5$ is found. The first AFM phase transition is clearly
observed as a maximum of $R(T)$ at about $T\approx4.9$ K. The inset
shows the data derivative and position of the dip in the derivative
is very close to $T_{N1}$ as determined from $C_{p}$. In Fig. \ref{fig:PrCoAl-resultsB}$(d)$,
we compare $C_{p}^{\text{Pr}}$ (left axis) with $d(T\chi_{c})/dT$
to pinpoint the values of $T_{N1}$and $T_{N2}$. The results obtained
from the peaks in resistivity, heat capacity and $d(T\chi_{c})/dT$
are in close agreement \citet{fisher_relation_1962}. A phase diagram,
determined from magnetization measurements (with $H\text{ }||\text{ }c$),
is presented in figure \ref{fig:PrCoAl-resultsB}$(e)$. 

\begin{figure}
\begin{centering}
\includegraphics[width=0.95\columnwidth]{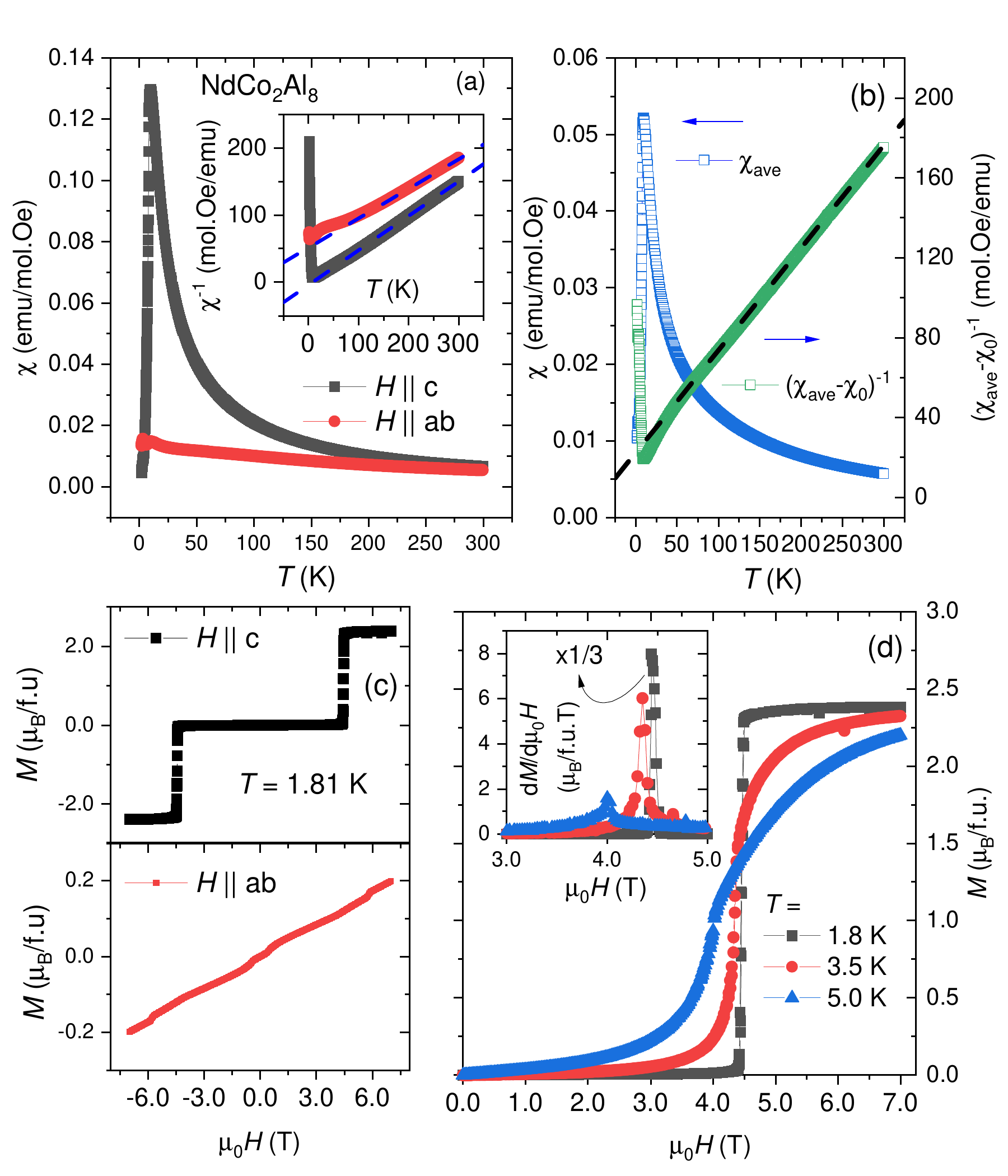}
\par\end{centering}
\caption{NdCo$_{2}$Al$_{8}$ magnetic properties. $(a)$ $\chi_{c}$ and $\chi_{ab}$
data obtained for $\mu_{0}H=0.1$ T. From $\chi_{c}$, we deduce an
AFM transition at $T_{N}=8.1$ K . In the inset, we present the inverse
of $\chi_{a}$ and $\chi_{ab}$ subtracted by $\chi_{0}$ . The dashed
blue lines represent the CW fitting (obtained for $T>100$ K) of the
data. $(b)$ Respectively on the left and right axis, $\chi_{\text{ave}}$
and $(\chi_{\text{ave}}-\chi_{0})^{-1}$ are presented. The dashed
black blue represent the CW fitting of the data (obtained
for $T>150$ K) $(c)$ Isothermal magnetization curves ($T=1.81$
K) obtained with $H\text{ }||\text{ }c$ (upper panel) and $H\text{ }||\text{ }ab$
(lower panel). $(d)$ Representative isothermal magnetization curves
for distinct temperatures obtained for $H\text{ }||\text{ }c$. The
data show clear spin-flops transitions from a low field AFM state
to a high field polarized metamagnetic state. The inset shows the
$dM/d\mu_{0}H$ derivatives from which we determined the critical
field for the field-induced metamagnetic transition. \protect\label{fig:NdCoAl-resultA}}
\end{figure}

\begin{figure}
\begin{centering}
\includegraphics[width=0.95\columnwidth]{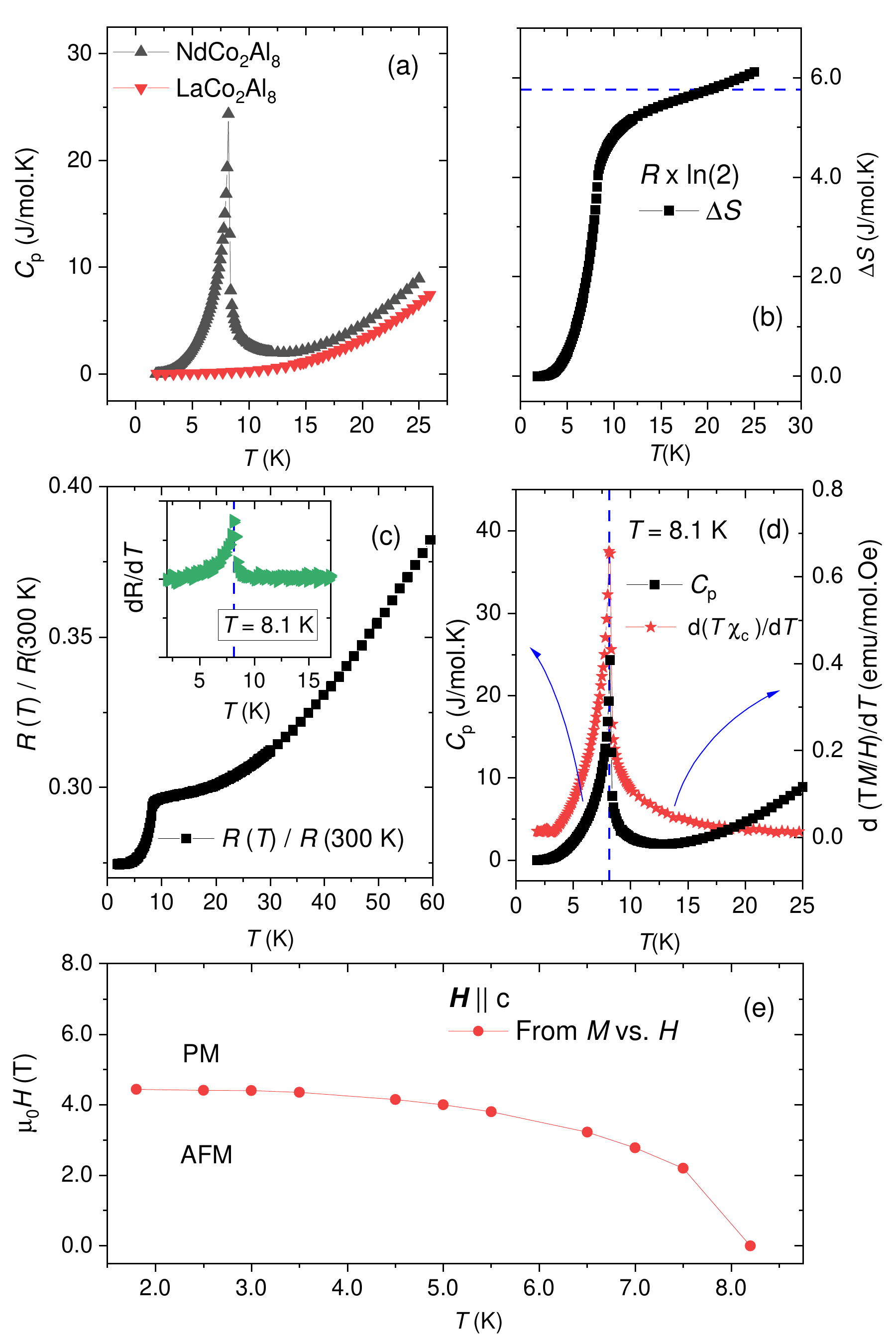}
\par\end{centering}
\caption{NdCo$_{2}$Al$_{8}$ specific heat and resistance data. $(a)$ NdCo$_{2}$Al$_{8}$
$C_{p}$ (black triangles) showing the AFM transitions. The LaCo$_{2}$Al$_{8}$
$C_{p}$ is presented as reference. $(b)$ NdCo$_{2}$Al$_{8}$ magnetic
entropy variation (see main text). $(c)$ Temperature
dependence of the normalized resistance $R(T)/R(300\text{ K})$. The
inset shows the resistance derivative, $dR/dT$ $(d)$ Comparison between $C_{p}$
and $d(T\chi_{c})/dT$ showing $T_{N}$ as determined from independent
measurements. $(e)$ NdCo$_{2}$Al$_{8}$ magnetic phase diagram deduced
from $M(H)$ ($H\text{ }||\text{ }c$) measurements. \protect\label{fig:NdCoAl-resultB}}
\end{figure}

Experiments of NdCo$_{2}$Al$_{8}$ were motivated by the magnetic
quantum critical behavior claimed to have been observed in isostructural NdFe$_{2}$Ga$_{8}$,
which seems to present the expected behavior of a three dimensional
spin density wave type of quantum critical point (QCP) \citep{wang_neutron_2022,wang_quantum_2021}.
Our findings show that the NdCo$_{2}$Al$_{8}$ case is simpler and
similar to that of the Pr-based material. Results are presented in
figures \ref{fig:NdCoAl-resultA}$(a)$-$(d)$ and \ref{fig:NdCoAl-resultB}$(a)$-$(e)$.
The NdCo$_{2}$Al$_{8}$ $\chi_{c}$ and $\chi_{ab}$ data are in
figure \ref{fig:NdCoAl-resultA}$(a)$ and show that the material
is also a strong easy-axis magnet which presents an AFM transition
at $T_{N}=8.1$ K. In the inset, we present the analysis of $\chi_{c}$ and $\chi_{ab}$,
obtaining $\theta_{c}=5(3)$ K and $\theta_{ab}=-125(7)$ K. 

To estimate the effective moments and size of the interactions, we
examine in figure \ref{fig:NdCoAl-resultA}$(b)$ $\chi_{\text{ave}}$
(left axis) and the inverse of $(\chi_{\text{ave}}-\chi_{0})^{-1}$
(right axis). Here, $\chi_{0}=2\times10^{-4}$ emu/mol.Oe representing
the LaCo$_{2}$Al$_{8}$ Pauli-like susceptibility. A CW constant
$(\theta_{\text{CW}}^{\text{Nd}}$) of $-43(2)$ K is obtained and
suggests AFM interactions. A frustration parameter $f\approx5$ is
deduced, half of what was obtained for the Pr-based material. The
obtained effective moment is $\mu_{\text{eff}}^{\text{Nd}}=3.95(5)$
$\mu_{\text{B }}$which compares well with the theoretical value for
the Nd$^{3+}$ cation ($3.61$ $\mu_{\text{B}}$). As in the Pr case,
the $\mu_{\text{eff}}^{\text{Nd}}$ value may indicate some contribution
from Co, but this is at the edge of our resolution, given the lack of a true polycrystalline average, as discussed above. 

Isothermal magnetization measurements at $T=1.81$ K, for $H\text{ }||\text{ }c$
(upper panel) and $H\text{ }||\text{ }ab$ (lower panel), are presented
in figure \ref{fig:NdCoAl-resultA}$(c)$. A metamagnetic transition
from a low field AFM state to a high field spin-polarized PM state
is observed at about $\approx4.5$ T for $H||c$. The value of the
saturated magnetization corresponds closely to $2.4$ $\mu_{\text{B}}/\text{f.u.}$, which could 
suggest a $\alpha\left\Vert \pm5/2\right\rangle \pm\beta\left\Vert \mp7/2\right\rangle $
doublet CF ground state with a large $\left\Vert \mp7/2\right\rangle $
component; on the other hand, a higher applied field could lead to another metamagnetic transition with a subsequent $\mu_{sat} \approx 3.3  \mu_B$. In the lower panel, it is shown that relative small fields
can generate a spin rotation transition. By comparing the values of
the magnetization at $1.81$ K and $\text{\ensuremath{\mu_{0}H=7}}$T
for the two field directions, the magnetic anisotropy in NdCo$_{2}$Al$_{8}$
is about $\approx12$.

In figure \ref{fig:NdCoAl-resultA}$(d)$, a series of representative
isothermal magnetizations ($H\text{ }||\text{ }c$) are presented
to examine the temperature dependence of the metamagnetic transition.
The field derivative of the magnetization ($dM/d\mu_{0}H$) are presented
in the inset of the figure. Again, we adopt the field at which the
derivatives peak as the critical field for the metamagnetic transition.

The LaCo$_{2}$Al$_{8}$ (reference data) and the NdCo$_{2}$Al$_{8}$
$C_{p}$ ($C_{p}^{\text{Nd}}$) are presented in figure \ref{fig:NdCoAl-resultB}$(a)$. 
The latter clearly shows the AFM transition at $T_{N}=8.1$ K . As in the PrCo$_{2}$Al$_{8}$ case, we adopt the LaCo$_{2}$Al$_{8}$
$C_{p}$ as a reference to estimate the NdCo$_{2}$Al$_{8}$ magnetic
heat capacity along with the calculation of a renormalization factor
$\Theta^{\text{Nd}}_{\text{D}}/\Theta^{\text{La}}_{\text{D}}$. As previously, we also investigated the normalization adopting different methods and temperature ranges for the fittings. The obtained factors lied in between $0.95(4)$ and $0.91(2)$. We normalized $C_{p}^{\text{Nd}}$ by $0.93$ (the average) to obtain the magnetic heat capacity and then the total magnetic entropy recovered for $T$ up to $25$ K. This is shown in figure \ref{fig:NdCoAl-resultA}$(b)$
and the result corresponds well to the ordering of a CF doublet ground
state associated with all the degrees of freedom in the sample. Indeed, being Nd$^{3+}$ a Kramer's cation, the CF scheme in its low symmetry environment should feature five doublets. 

The NdCo$_{2}$Al$_{8}$ normalized resistance is shown in figure
\ref{fig:NdCoAl-resultB}$(c)$. A sharp drop of $R(T)$ at about
$8$ K axis marks the AFM transition. The $RRR$  is also close
to $\approx5$ as for the other $R$-based materials. In the inset,
we show the derivative of the normalized resistance and a peak is
observed at $T=8.1$ K, marking the AFM transition. In \ref{fig:NdCoAl-resultB}$(d)$,
$C_{p}^{\text{Nd}}$ (left axis) and $d(T\chi_{c})/dT$ are compared
to pinpoint the value of $T_{N}$, which is confirmed as $8.1$ K.
We summarize our findings in the phase diagram in figure \ref{fig:NdCoAl-resultB}$(e)$. 


\subsection{SmCo$_{2}$Al$_{8}$}

\begin{figure}
\begin{centering}
\includegraphics[width=0.95\columnwidth]{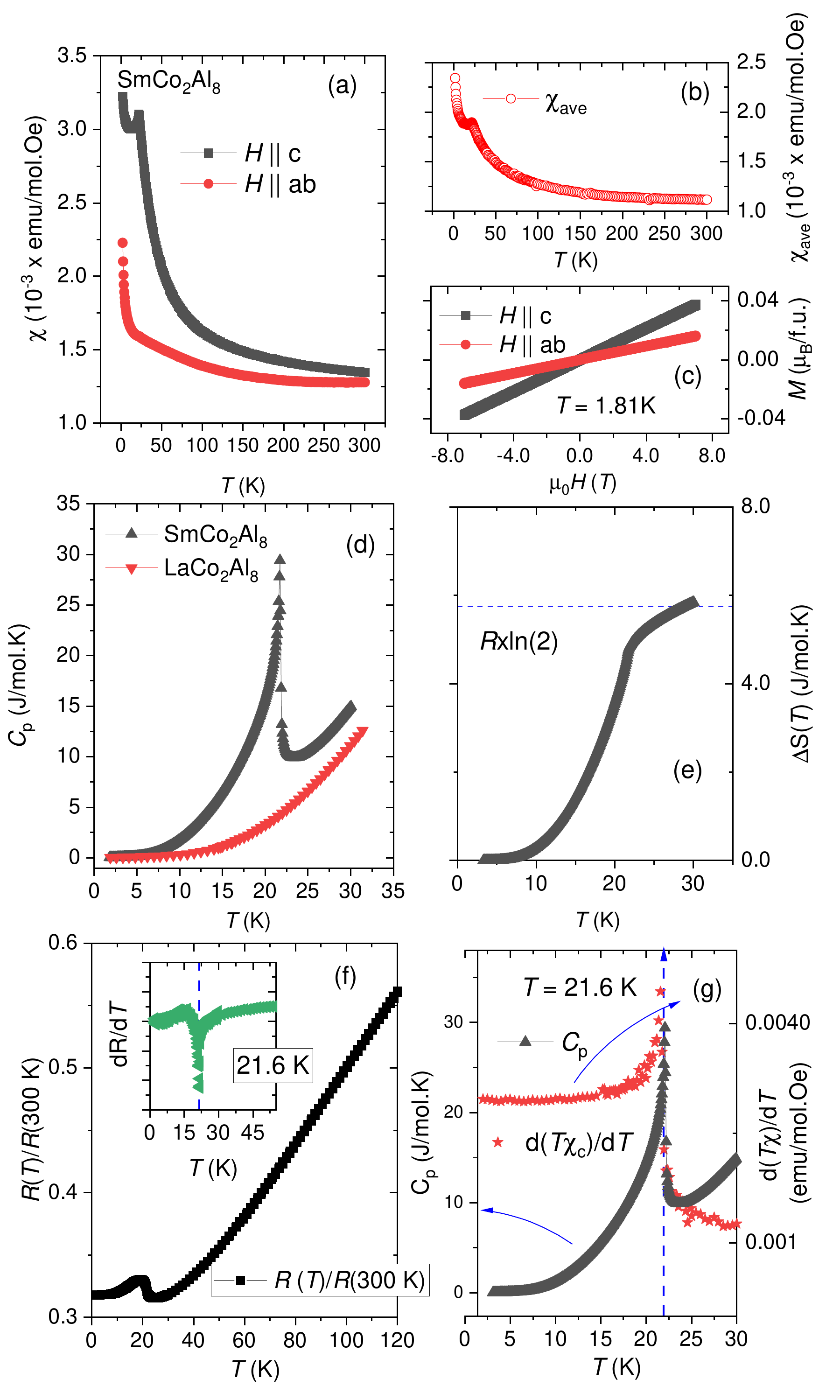}
\par\end{centering}
\caption{SmCo$_{2}$Al$_{8}$ magnetic, thermal and transport properties. $(a)$
$\chi_{c}$ and $\chi_{ab}$ data obtained for $\mu_{0}H=0.1$ T.
An AFM transition is suggested at $T_{N}=21.6$ K for measurements
along the $c$-axis. $(b)$ $\chi_{\text{ave}}$ is presented. CW
behavior is not observed in the investigated $T$-interval $(c)$
Isothermal magnetization curves ($T=1.81$ K) obtained with $H\text{ }||\text{ }c$
and $H\text{ }||\text{ }ab$ $(d)$ SmCo$_{2}$Al$_{8}$ $C_{p}$
(black triangles) showing the AFM transition at $T_{N}=21.6$ K as
suggested by $\chi$ measurements. The LaCo$_{2}$Al$_{8}$ $C_{p}$
(red triangles) is presented as reference. $(e)$ SmCo$_{2}$Al$_{8}$
magnetic entropy variation (see main text). $(f)$ Temperature
dependence of the normalized resistance $R(T)/R(300\text{ K})$. The
inset shows the resistance derivative, $dR/dT$. $(g)$ Comparison between $C_{p}$
and $d(T\text{\ensuremath{\chi_{c}}})/dT$ determining $T_{N}$ from
independent measurements. \protect\label{fig:SmCoAl-results}}
\end{figure}

The SmCo$_{2}$Al$_{8}$ properties are presented in figures \ref{fig:SmCoAl-results}$(a)$-$(g)$.
In figure \ref{fig:SmCoAl-results}$(a)$ $\chi_{c}$ and $\chi_{ab}$
data characterize SmCo$_{2}$Al$_{8}$ as an easy-axis magnet with
$T_{N}\approx21.6$ K. In \ref{fig:SmCoAl-results}$(b)$, we present
$\chi_{\text{ave}}$. The overall paramagnetic response is
small and CW behavior is not observed. This is common for Sm-based intermetallic compounds where the first excited Hund's rule multiplet is low enough so as to contaminate the ground-state multiplet response.

SmCo$_{2}$Al$_{8}$ was previously reported as a Pauli paramagnet,
wherein Sm cations assume a non-magnetic $2+$ valence \citep{watkins-curry_strategic_2015}.
A careful inspection of the data is thus required. We first investigate
the low-$T$ upturn of the data. We present in figure \ref{fig:NdCoAl-resultA}$(c)$
isothermal magnetization measurements measured at $T=1.81$ K, with
$H\text{ }||\text{ }c$ and $H\text{ }||\text{ }ab$. The response
is small for both field directions and describe the finite response
of the AFM state to an applied field. No ferromagnetic component is
observed and a paramagnetic component (from any impurity phase) would
have a larger response. 

The LaCo$_{2}$Al$_{8}$ and SmCo$_{2}$Al$_{8}$ $C_{p}$ are presented
in figure \ref{fig:SmCoAl-results}$(d)$. The phase transition suggested
in figure \ref{fig:SmCoAl-results}$(a)$ is clearly present in the $C_{p}^{\text{Sm}}$
measurements. To normalize the phonon contribution, we modeled 
$C_{p}^{\text{La}}$ and $C_{p}^{\text{Sm}}$ in the $T$-interval $28<T<75$, which is well above the transition temperature, taking into account Debye and optical phonons. Indeed, this $T$-range lies outside the validity of the $T^{3}$ law. The obtained renormalization factor is $0.86(3)$. We then obtain
the SmCo$_{2}$Al$_{8}$ magnetic heat capacity. It is observed that at low-$T$ it becomes slightly negative, certainly because of the uncertainties related with the phonon subtraction. We then integrate it in the interval $3.3<T<30$ K, where it is strictly positive,  to find the total recovered magnetic entropy up to $30$ K. 
Results are shown in figure \ref{fig:SmCoAl-results}$(e)$. As can
be observed, $R\times\text{ln}(2)$ is obtained at about $T\approx29$
K. This result is certainly underestimating the magnetic entropy  but, nevertheless, shows that the transition here observed corresponds to the ordering of a well localized ground state CF doublet and suggests the absence of Sm mixed-valence. Moreover, the results give further confidence that
the SmCo$_{2}$Al$_{8}$ magnetic response stems from the whole of
the sample and not from an impurity. 

The SmCo$_{2}$Al$_{8}$ transport properties are examined in \ref{fig:SmCoAl-results}$(f)$,
where the normalized resistance as a function of $T$ is presented.
The obtained $RRR$ is close to $\approx4$. A broad peak in $R(T)$
is observed about $T\approx21$ K and can be ascribed to the AFM transition.
In the inset, we present the resistance derivative and the peak position
in the derivative is identified at $T=21.6$ K. Furthermore, results
for $R(T)$ do not support Sm mixed-valence. For a more precise determination
of $T_{N}$ , we show in Fig.\ref{fig:SmCoAl-results}$(f)$ $C_{p}^{\text{Nd}}$
(left axis) and $d(T\chi_{c})/dT$ (right axis). The AFM transition
temperature $T_{N}=21.6$ K is confirmed. 

\section{Summary and Conclusions }

We investigated the anisotropic magnetic and transport properties
and thermal properties of $R$Co$_{2}$Al$_{8}$ single crystals ($R = $ La, Ce, Pr, Nd, and Sm). LaCo$_{2}$Al$_{8}$ is a Pauli paramagnet with a resolvable anisotropy in the temperature dependent resistivity. CeCo$_{2}$Al$_{8}$ is a Kondo lattice compound that presents no magnetic order
down to $1.81$ K. Anisotropic Kondo coherence temperatures $T_{\text{K,}c}^{*}\approx68$
K and $T_{\text{K,}ab}^{*}=46$ K were estimated from the broad peaks
of the Ce $4$f contribution to $\rho_{c}$ and $\rho_{ab}$, respectively.
A thermodynamic Kondo temperature was estimated from $C_{p}$ and
amounts to $T_{_{K}}= 36-70$ K, depending on the method that is adopted to estimate it. 

Based upon these results, we concluded that an anisotropic Kondo state
is formed in CeCo$_{2}$Al$_{8}$ in a way that coherent Kondo scattering
first sets in along the $c$-axis and then along the $ab$ plane.
This anisotropy likely stems from the conduction electrons anisotropy,
measured for both the LaCo$_{2}$Al$_{8}$ and CeCo$_{2}$Al$_{8}$ compounds. In
a sense, in the temperature interval in between $T_{\text{K,}c}^{*}$
and $T_{\text{K,}ab}^{*}$, one finds a situation that is similar
to what is observed in CeCo$_{2}$Ga$_{8}$ below $17$ K, and was
termed an axial Kondo chain. 

The PrCo$_{2}$Al$_{8}$ and NdCo$_{2}$Al$_{8}$ materials are easy
axis AFM materials for which the moments are aligned along the $c$-axis. For PrCo$_{2}$Al$_{8}$,
two AFM transitions are observed at zero applied field. The transitions
merge into a single transition at low-$T$. It is safe to assert that in PrCo$_{2}$Al$_{8}$, the CF ground state
is a $\left\Vert4\right\rangle $ singlet making PrCo$_{2}$Al$_{8}$
an Ising-like system. 

The SmCo$_{2}$Al$_{8}$ material is an AFM material, but no information
about the single ion physics could be extracted, since a CW behavior
was not observed. For all AFM materials, $T_{N}$ was confirmed by
heat capacity, magnetization and resistivity measurements. 

\section{Acknowledgments}

This work was supported by the U.S. Department of Energy, Office of
Science, Basic Energy Sciences, Materials Science and Engineering.
Ames National Laboratory is operated for the USDOE by Iowa State University
under Contract No. DE-AC02-07CH11358. F.G. and C.A acknowledge, respectively,
financial support from the University of Sao Paulo and UNICAMP (sabbatical
leave). Ryan Mackenzie is acknowledged for fruitful discussions during
the draft of the manuscript.

\bibliography{2023RCo2Al8_references}

\end{document}